\documentclass[conference]{IEEEtran}
\IEEEoverridecommandlockouts
\usepackage{amsmath,amssymb,amsfonts}

\usepackage{graphicx}
\usepackage{textcomp}
\usepackage{xcolor}
\def\BibTeX{{\rm B\kern-.05em{\sc i\kern-.025em b}\kern-.08em
    T\kern-.1667em\lower.7ex\hbox{E}\kern-.125emX}}

\usepackage{mathptmx} %
\usepackage{adjustbox}
\usepackage{paralist}
\usepackage{verbatim}
\usepackage{fancyhdr}
\usepackage[normalem]{ulem}
\usepackage[hyphens]{url}
\usepackage[sort,nocompress]{cite}
\usepackage[final]{microtype}
\usepackage{listings}
\usepackage{ntheorem,lipsum}
\usepackage{adjustbox}
\usepackage{amsmath}
\usepackage{amsmath}
\usepackage{longtable}
\theorembodyfont{\upshape}

\usepackage{graphicx}
\usepackage{textcomp}
\usepackage{xcolor}
\usepackage{setspace}
\usepackage{color}
\usepackage{makecell}
\usepackage{adjustbox}

\usepackage{titlesec}

\usepackage{caption} 
\newcommand{\ignore}[1]{}

  {\colorbox{yellow}}%
  {}
\def\BibTeX{{\rm B\kern-.05em{\sc i\kern-.025em b}\kern-.08em
    T\kern-.1667em\lower.7ex\hbox{E}\kern-.125emX}}
\makeatletter
\newcommand{\linebreakand}{%
  \end{@IEEEauthorhalign}
  \hfill\mbox{}\par
  \mbox{}\hfill\begin{@IEEEauthorhalign}
}
\makeatother
 
\begin{document}
\onecolumn
\thispagestyle{empty}
\vspace*{1cm}
{\LARGE\bfseries PerSpectron: Detecting Invariant Footprints of Microarchitectural Attacks with Perceptron\par}
\vspace{1cm}
{\Large Author manuscript\par}
\vspace{0.6cm}
2020 53rd Annual IEEE/ACM International Symposium on Microarchitecture (MICRO), pp. 1124-1137 (2020)\par
\vspace{0.4cm}
DOI: \url{https://doi.org/10.1109/MICRO50266.2020.00093}\par
\vfill
{\small \textcopyright\ 2020 IEEE. Personal use of this material is permitted. Permission from IEEE must be obtained for all other uses, in any current or future media, including reprinting/republishing this material for advertising or promotional purposes, creating new collective works, for resale or redistribution to servers or lists, or reuse of any copyrighted component of this work in other works.\par}
\clearpage
\twocolumn
\setcounter{page}{1}

\title{PerSpectron: Detecting Invariant  Footprints of Microarchitectural Attacks with Perceptron\\

}

\author{\IEEEauthorblockN{Samira Mirbagher-Ajorpaz}
\IEEEauthorblockA{\textit{Computer Science and Engineering} \\
\textit{Texas A\&M University}\\
College Station, USA \\
samiramir@tamu.edu}
\and
\IEEEauthorblockN{Gilles Pokam}
\IEEEauthorblockA{
\textit{Intel Labs}\\
Santa Clara, USA \\
gilles.a.pokam@intel.com}
\and
\IEEEauthorblockN{Esmaeil Mohammadian-Koruyeh}
\IEEEauthorblockA{\textit{Computer Science and Engineering} \\
\textit{University of California, Riverside}\\
Riverside, USA \\
emoha004@ucr.edu}
  \linebreakand %
\IEEEauthorblockN{Elba Garza}
\IEEEauthorblockA{\textit{Computer Science and Engineering} \\
\textit{Texas A\&M University}\\
College Station, USA \\
elba@tamu.edu}
\and
\IEEEauthorblockN{Nael Abu-Ghazaleh}
\IEEEauthorblockA{\textit{Computer Science and Engineering} \\
\textit{University of California, Riverside}\\
Riverside, USA \\
naelag@ucr.edu}
\and
\IEEEauthorblockN{Daniel A. Jiménez}
\IEEEauthorblockA{\textit{Computer Science and Engineering} \\
\textit{Texas A\&M University}\\
College Station, USA \\
djimenez@acm.org}
}

\maketitle
\begin{abstract}
Detecting microarchitectural attacks is critical given their proliferation in 
recent years. Many of these attacks exhibit intrinsic behaviors essential to the 
nature of their operation, such as creating contention or misspeculation. This 
study systematically investigates the microarchitectural footprints 
of hardware-based attacks and shows how they can be detected and classified using 
an efficient hardware predictor.
We present a methodology to use correlated microarchitectural statistics 
to design a hardware-based neural predictor capable of detecting and classifying 
microarchitectural attacks before data is leaked. Once a potential attack is 
detected, it can be proactively mitigated by triggering appropriate 
countermeasures.

Our hardware-based detector, {\em PerSpectron}, uses perceptron learning to
identify and classify attacks. Perceptron-based prediction has been successfully 
used in branch prediction and other hardware-based applications.
PerSpectron has minimal performance overhead. The statistics being monitored 
have similar overhead to already existing performance monitoring counters.
Additionally, PerSpectron operates outside the processor's critical paths,
offering security without added computation delay. Our system achieves a usable 
detection rate for detecting attacks such as SpectreV1, SpectreV2, SpectreRSB, Meltdown, 
breakingKSLR, Flush+Flush, Flush+Reload, Prime+Probe as well as cache-attack 
calibration programs.  We also believe that the large number of diverse 
microarchitectural features offers both evasion resilience and interpretability---features 
not present in previous hardware security detectors. We detect these attacks early 
enough to avoid any data leakage, unlike previous work that triggers countermeasures 
only after data has been exposed.
\end{abstract}
\section{Introduction} 

The number of known microarchitectural security vulnerabilities caused by speculative 
execution has increased sharply in recent years~\cite{canella2018systematic}.
More recently, microarchitectural data sampling (MDS) vulnerabilities 
have been introduced: RIDL\cite{ridl}, Fallout\cite{canella2019Fallout} and 
LVI~\cite{vanbulck2020lvi} exploit these vulnerabilities to leak data from
internal CPU buffers. Also, the new CacheOut\cite{cacheOut} attack adds the ability 
to select which data to leak from the L1 cache to the MDS-based attack. 

Various components in a modern processor pipeline are susceptible to 
attacks~\cite{canella2018systematic}, exposing 
many side-channels. A large body of work has been devoted to identifying each 
of these attacks, with industry responding with appropriate
patches. However, this is a cat-and-mouse game as new attacks continue to 
appear at a regular cadence.

Prior work on detecting side-channel attacks relies on querying hardware 
performance counters via software to reveal malevolent behavior and prevent 
data from being leaked~\cite{cacheBasedDetection2016Chiappetta, 
  CloudRadar2016, BlackHatFogh, payer2016hexpads, ICCAD2015Wang, 
  Duppel2013Zhang,mushtaq:cel-01824512}. 
Unfortunately, relying on performance counters poses practical problems 
that can hurt detection and performance. First, the number of events available 
to performance counters is limited, {\em e.g.} on recent Intel processors, 
performance counters can only monitor up to 4 events at a time. Thus, detection 
techniques that require monitoring a rich feature space must 
multiplex counters, potentially degrading accuracy, providing opportunities 
for evasion, and missing opportunities to detect an attack. Second, because performance 
counters are accessible in software, an attacker may be able to access them and modulate its 
attack to evade detection. Third, the sampling rate of performance counters is 
low enough to allow an attacker to adjust the bandwidth of an attack to fit within the 
sampling interval of a software based detector.

We propose an alternative approach: a first line of defense that protects 
systems against broadly-defined microarchitectural attacks by leveraging 
a distinct {\em microarchitectural footprint} of each attack to detect and classify 
these attacks-in-progress. This technique enables countermeasures to be deployed 
proactively, before the attack can be successful. 
There is evidence that microarchitectural attacks are being used to hide other, 
more traditional attacks as well, so it is important for security solutions 
to be able to detect attacks in the speculative execution feature 
space~\cite{wampler-19}.  

Our hardware classifier can capture the signature of an attack in
a much richer feature space, allowing higher classification accuracy.  
Security guarantees are stronger in systems that rely
on microarchitectural statistics and are monitored in hardware. 
The detector cannot be disabled by software even if the kernel is compromised, 
which is important for threat scenarios where a compromised kernel may 
be attempting to compromise a secure enclave~\cite{chen-18,wang-17}.  Lastly, our 
hardware-based detector has a sampling frequency that makes it impossible for an 
attacker to time its attack to the sampling interval. 

Our proposed approach to detecting microarchitectural attacks moves the 
detection of active attacks to hardware from software, allowing the 
predictor to efficiently use and monitor a large set 
of microarchitectural features that is not limited to the commit state and 
includes speculative instructions. There has been prior work on detecting 
malware in hardware~\cite{Malware2015, ensembleRaid2015,kazdagli-16,RHMD2017}.

However those works were (1) not specific to microarchitectural attacks; 
and (2) primarily look at features related to committed state such as instruction 
mixes or memory access distribution. This is important because the signature for 
microarchitectural attacks is different than the signature of malware. We compare 
PerSpectron to these and other works in Section~\ref{compareMLs}.

Designing such a system in hardware poses multiple challenges.
First, we need a way to select a set of features to use as input to
the detector. The features should be strongly correlated to known  
microarchitectural attacks. Identifying features that correlate well with an active 
microarchitectural attack is challenging, as it requires understanding 
the subtle interactions between various and plentiful pipeline stages across 
different processor components.  It is also important that these features 
be discriminatory; otherwise, attackers may be able to evade them. 
Worse, normal programs may lead to false positives, reducing the effectiveness 
of the defense.

We propose a perceptron-based algorithm that streamlines the feature selection 
process. From the 1159 microarchitectural features available in our hardware simulator, 
our algorithm selects 106 features that show the strongest correlation to a range of 
microarchitectural attacks, including SpectreV1, SpectreRSB, 
Meltdown, breakingKSLR, Flush+Flush, Flush+Reload, Prime+Probe, and 
Calibration. The selected features are then used as inputs to a hardware 
predictor for microarchitectural attack detection and classification.
https://www.overleaf.com/project/5f444c7784c8540001be8e03The attacks that we examine are complex, yet identifiable. 

The second challenge is designing the mechanism to predict and classify 
microarchitectural attacks rapidly and with both high precision and recall. 
The artificial neural networks that are used currently to recognize 
microarchitectural attacks in software typically use multiple layers of learned 
feature detectors that produce scalar outputs. By contrast, we use 
replicated perceptron detectors in hardware with overlapping features. 
We know that deep neural networks learn complicated features using multiple 
hidden layers. We show how replicated-perceptron detectors with a large set 
of low-level features can be used to detect complicated features related to 
microarchitectural attacks without hidden layers. We argue that this is a 
more promising way of dealing with variations in speculative attacks 
and cache attacks than the methods currently employed.

With any machine learning based detection solution, there are concerns about the 
ability of the attackers to evade detection by modifying their behavior to fool the 
detector~\cite{RHMD2017}. We believe that the nature of microarchitectural attacks, 
and the fact that most of them are timing sensitive, substantially limit the ability 
of the attackers to evade detection. Moreover, we believe that solutions such as 
RHMD~\cite{RHMD2017} where multiple detectors are used and invoked stochastically can 
substantially mitigate the threat of evasion. Another consideration with hardware 
detection is the threat of false positives: our detectors experience a very low false 
positive rate.  Moreover, we view them as a first line of defense that simply alerts 
the operating system to enable the invocation of subsequent mechanisms to isolate a 
suspicious process or to monitor it more closely. 

This paper makes the following contributions:
\begin{enumerate}
 \item A novel use of perceptron learning to detect and classify a broad range 
 of microarchitectural attacks. We introduce new replicated perceptron-based algorithms 
 for the selection of highly correlated invariant features of microarchitectural 
 attacks across, the different pipeline stages, most relevant for attack classification. 
 \item A comprehensive analysis of microarchitectural features that indicate 
 footprints of an attack. We show using the proposed detection and 
 classification algorithms, that valuable insights pertaining to the 
 properties of microarchitectural attacks can be gleaned from a systematic 
 analysis of the weights of the features in the perceptron design.
 \item A hardware design of the proposed detection system and a fast 
 classification system with low area and performance overhead. We provide proof 
 that it is robust to even tweaked variations of considered attacks, including 
 one of Flush+Flush, which prior work could not detect in hardware.
 \end{enumerate} 

\section{Background and Motivation}\label{background}
The following two subsections explain the attacks considered in our work. 
We categorize them as (1) cache side-channel attacks, which exploit 
timing differences caused by the lower latency of CPU caches 
(compared to physical memory), and (2) speculative attacks, 
which exploit microarchitectural vulnerabilities and leak the data 
using the above-described side-channel attacks. 

In particular, for cache side-channel attacks, we examine Prime+Probe, 
Flush+Reload, and Flush+Flush attacks. For speculative (also known as 
transient execution) attacks, Spectre Variant1 (SpectreV1), Meltdown, 
and SpectreRSB. Our training set is limited to these six attacks.\footnote{For 
practical reasons, our data set does not include all known microarchitectural 
attacks  ({\em e.g}, some variants of speculation attacks, side channels on 
other CPU structures, and RowHammer attacks); however, we believe 
PerSpectron can be trained to detect additional attacks.}

We then describe the perceptron, a small, single-layered learning network
that is cheaply implementable in hardware and is the foundational computational 
structure of our work.

\subsection{Cache attacks}
In a Prime+Probe attack~\cite{PrimeProbe2015last}, the attacker primes the cache 
by filling cache sets with its own code or data, then waits for a set time while 
the victim executes. Finally, the attacker probes the cache by executing and 
measuring the time to load each set of the primed data or code. If the victim has 
accessed some cache sets, it will have evicted some of the attacker's cache 
lines, which the attacker observes via increased access latencies for those lines.

In a Flush+Reload attack~\cite{FlushReload2014Yarom}, the attacker and the 
victim share some pages in their address space, {\em e.g.} via shared libraries. 
First, the monitored data is flushed from the cache hierarchy. Second, 
the attacker waits to allow the victim time to access this data. Third, 
the attacker reloads the memory line, measuring the time to load it. If during 
the waiting phase the victim accesses the data, the line will be present in 
the cache and the reload operation will take a short time, exposing the 
victim's access behavior. If the victim has not accessed the memory 
line, the attacker will observe a longer access time. 

Unlike the above attacks, a Flush+Flush attack ~\cite{GrussFlushFlush} does 
not make any memory accesses and causes no cache misses from the attacker process. 
Instead, this attack relies on the execution time of the flush instruction: 
if an address is present in the cache, flushing takes more time. 
Flush+Flush attacks are stealthy~\cite{GrussFlushFlush}, {\em i.e.} the spy process 
cannot be detected based on cache hits and misses detection 
mechanisms\cite{GrussFlushFlush}. Because Flush+Flush conducts neither
caches accesses nor causes misses, it cannot be detected by Cyclone~\cite{cyclone2019},
a state-of-the-art side channel detection solution.  

\subsection{Speculative attacks}
\textit{Spectre Variant1}~\cite{spectre}, or SpectreV1, exploits branch predictors 
used to predict the direction of conditional branches. The attacker mis-trains 
the branch predictor to bypass the boundary check of a data structure and 
speculatively access secret data out of bounds. As a result, the CPU 
speculatively brings indexing data into the cache that would not have 
been loaded otherwise. The attacker uses the secret data to index 
into its own address space and uses a cache side-channel (such as 
Flush+Reload) to leak the data. 

The \textit{Meltdown} attack~\cite{meltdown} relies on the fact that permission
checks are performed late in the execution pipeline, and a fault is generated only 
for a committed instruction. From the time an instruction is marked for permission
exception at the beginning of the pipeline, to the time that an exception can be raised, 
if the attacker makes a (normally unauthorized) memory access, the secret data can be 
cached and leaked using a side-channel.

\textit{SpectreRSB}~\cite{koruyeh2018spectre} is another Spectre attack variant
that exploits the return stack buffer (RSB), a predictor unit to predict the target of 
return instructions. In this attack, the attacker pollutes the RSB with an unmatched 
call/return pair, which redirects the speculative control flow of the program to 
a malicious Spectre-like setup.

\subsection{Perceptron}\label{perceptron}
We address software limitations by moving the detection of
attacks to hardware, allowing the selection of features from a 
larger set of events without incurring performance overhead. 
Neural network models used in current work feature deep multi-layered 
networks ({\em e.g.} RNN) that are not easily amenable to hardware due 
to design and runtime complexity. We show that using a simple 
single-layered perceptron can provide a readily implementable solution.
Perceptron learning has shown to be implementable in hardware for various 
applications including branch prediction, prefetching, replacement policies, and CPU
adaptation~\cite{intelISCA2019}. Recent microarchitectures from Oracle~\cite{SPARCT4}, 
AMD ({\em e.g.} Bobcat, Jaguar, Piledriver, Zen, etc.), and Samsung~\cite{Mongoose,M3} 
are documented as featuring perceptron-based branch predictors.

We adapt the single-layer perceptron model used by prior work for detecting and classifying
side-channel attacks. 
A perceptron is a vector of weights that records correlations between an input
vector and a target value. It can be used to classify inputs into one of two
classes~\cite{RosenblattPerceptron}. We construct a quantized desired response $d(n)$ to a perceptron:
\[
  d(n)=\begin{cases}
    +1, & \text{if $x(n)$ belongs to malicious class}.\\
    -1, & \text{if $x(n)$  belongs to benign class}.
  \end{cases}
\]
The perceptron computes the weighted sum of the input patterns $x(n)$ 
and compares it to a threshold value.
If the sum exceeds the threshold, the output of the perceptron is +1, otherwise, is it -1. 
The perceptron weights are updated after the desired outcome $d(n)$ of the predicted
event is known. If the prediction was correct then the weights remain unchanged. 
Otherwise, the inputs are used to update the corresponding weights:\\
$w(n+1)=w(n)+\mu[d(n)-y(n)]x(n)$
where $\mu$ is the learning-rate parameter and the difference $d(n)-y(n)$ is the 
error signal.

Perceptrons are a natural choice for building a microarchitectural attack
detector because they can be efficiently implemented in hardware. Other forms
of neural networks such as those trained by back-propagation, and other forms
of machine learning such as decision trees, are less attractive because of
excessive implementation costs. 
\section{Threat Model}

The main goal of \textit{PerSpectron} is to detect cache-based side channel
attacks and speculative attacks and prevent the attacker from leaking the memory 
contents of a victim program that is otherwise not accessible to the attackers.
In the attack scenario, we assume a strong adversary model where the 
attacker may execute any code with the privileges of a normal user. Our defense also 
applies for attacks that target secure enclaves, where the attacker may even 
have operating systems privileges and use them to attack the enclave 
using microarchitectural attacks, although the appropriate action upon the 
detection of an attack may be different (for example, we may tear down 
the enclave). Note that our detectors are specialized to microarchitectural 
attacks; traditional software vulnerabilities such as memory corruption bugs, or 
physical side channel attacks, are different threat models and out of scope 
for our defense.

\section{PerSpectron Overview}\label{alg}
\subsection{General Approach}
\begin{figure}[!h]
  \centering 

  \begin{minipage}[t]{0.5\textwidth} \includegraphics[width=0.48\textwidth]{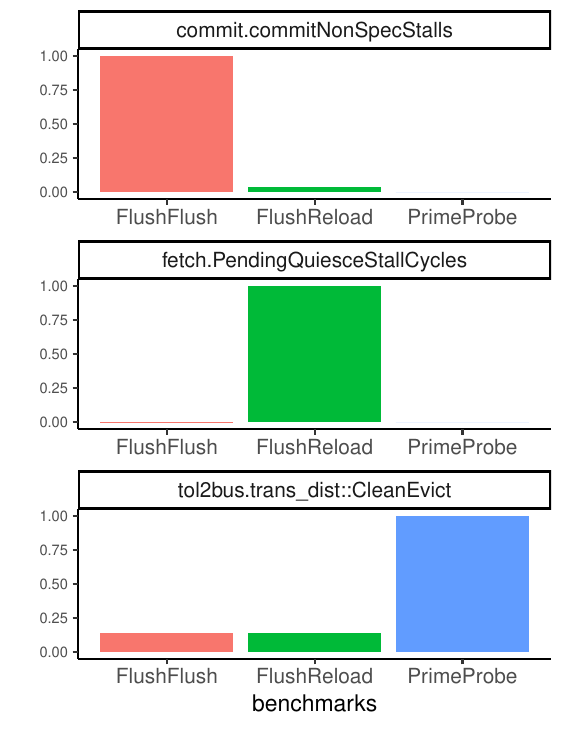}\hfill \begin{minipage}[t]{0.47\textwidth} 
  \includegraphics[width=\textwidth]{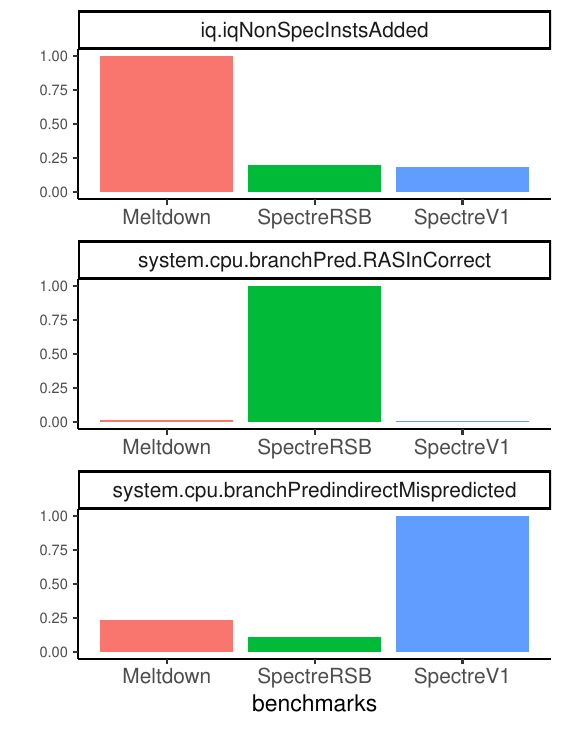} \end{minipage}
\vfill \begin{minipage}[t]{0.50\textwidth} \label{fig:phenom1}\end{minipage}\hfill \begin{minipage}[t]{0.47\textwidth}
  \end{minipage}  
  \end{minipage} 
    \caption[Caption for LOF]%
    {Information hops between input dimensions. To detect different variants of attacks we need \textit{viewpoint invariant features} and \textit{replicated detectors}. }
\label{fig:phenomm}
\end{figure}

In microarchitectural attacks, information about suspicious activity comes from 
different parts of the processor. We observe this for different attack variants,
as well as attempts to evade detection. For example, in SpectreV1, the attacker 
trains the branch prediction unit, while other variants use the Branch Target 
Buffer (BTB) or the return stack buffer (RSB). Figure~\ref{fig:phenomm} 
shows that information on the activity of these attacks hops between 
input domains. 

We can see that while SpectreRSB does not have a high 
\textit{branchPredindirectMispredicted} value, it has a high \textit{RASIncorrects} 
value. Meltdown cannot be predicted with a high \textit{SquashedLoads} value, but 
it causes high \textit{NonSpecInstrsAdded} to the IQ. Different variants of attacks 
present a problem similar to the \textit{viewpoint} problem of 
image recognition that standard learning models struggle 
with~\cite{Hinton1985shape,Hinton2018Capsule}.
The same is true for cache attacks\footnote{While Flush+Flush does not cause high
\textit{PendingQuiesceStallCycles}, it causes abnormal \textit{commitNonSpecStalls}. 
Also, while Flush+Reload does not cause abnormal \textit{commitNonSpecStalls}, it 
causes an abnormal number of \textit{PendingQuiesceStallCycles}. While Prime+Probe 
does not cause abnormally low \textit{tol2bus.transdist.CleanEvict} or 
\textit{PendingQuiesceStallCycles} values, it causes abnormal 
\textit{tol2bus.transdist.CleanEvict} values(See Figure~\ref{fig:phenomm}).}.
Each variant of an attack or evasion presents some common features and some unique 
compared to others, similar to the way an image does when it is rotated.  
We need to address this problem to detect different attack variants.

Importantly, common evasion techniques are unlikely to be effective against 
detecting distinct architectural footprints. For example, using polymorphism to 
produce different binaries, which can be effective against signature detection 
or features that look for specific instruction distributions, is unlikely to 
evade PerSpectron. Since microarchitectural attacks have distinctive time-sensitive 
microarchitectural footprints ({\em e.g.}., misspeculation contention), evasion is 
difficult. For example, if the attack slows to the point where it cannot execute 
within the misspeculation window, it is no longer effective. Moreover, we believe 
that techniques for evasion-resilience~\cite{RHMD2017} used by other hardware 
detectors can also be applied to PerSpectron to further improve its resilience 
to evasion. We will explore this idea in future work.

We know that information about attacks moves around the processor and shows up in 
different places similar to an image being rotated. We should detect these moving 
signatures in any component of the processor where they appear. Thus, we use the 
replicated features approach~\cite{lecun1999object}, an established solution for this 
type of problem. The idea is that if a feature vector was useful in detecting one target, 
it is likely that a similar feature detector with different positions in the input 
space can detect the replicated information, amplifying the signal and attenuating the 
noise.

Local features yield the well-known advantages of replicated features. 
Convolutional networks force the extraction of local features by restricting 
the receptive of hidden units to be local\cite{lecun1995convolutional}. 
Pipeline components already exhibit this property. Table~\ref{tab:replicate} shows 
a subset of replicated features from different pipeline stages that we found 
equally correlated in classifying attacks. 

We do not disregard mutually correlated features of different types, 
but rather we use them to our advantage. Hinton~\cite{Hinton1985shape} 
finds that using the entire space of possible instantiation parameters in the 
training set allowed the use of a simpler architecture, which could efficiently 
handle more complex images. In addition, no hidden layer was necessary---the 
mapping from the features' instantiation parameters to the object's 
instantiation parameters became linear. 

Our solution to this viewpoint problem starts with extracting a large, redundant 
set of features across each pipeline stage that are invariant under transformation and 
evasion similar to the invariant feature approach. Ullman~\cite{ullman2004recognition} 
observed that a large set of invariant features is sufficient to solve the problem.
By using the entire space of possible features in the training set we can efficiently 
detect multiple variations of attacks and evasions without needing to represent 
the relationships between features directly because other features capture 
them. PerSpectron applies the same method in a single layer. Using detailed 
microarchitectural features in hardware enables us to apply the invariant 
feature approach. 

An example of an invariant feature is the effect of misses and stalls 
in the Fetch stage. The squashed cycles in each stage, all the ROB, IQ, and 
Register full events, undone maps in the Rename stage, and memory order 
violation in the IEW stage propagate back to the Fetch stage. The relationship 
between these events' Fetch is not a simple cumulative function in an out-of-order 
processor. However, features such as \textit{fetch.MiscStallCycle} capture the 
relationship. By including overlapping and redundant features, one feature tells 
us how other features are related. This solves the domain-hopping problem caused by 
different variants of microarchitectural attacks and also makes our detector 
more difficult to evade. 

\subsection{Feature selection}
In this section, we explain extracting the invariant features and building the 
replicated perceptron detectors. 

The first step of the selection procedure derives, for each of the 1159 features, a 
measure of mutual information. We use the Pearson correlation coefficient, a 
number between -1 and 1 that indicates the extent to which two variables are linearly 
related. We then create a correlation matrix and, based on the values in it, we group 
features ${f_{1}\dots f_{m}}$ that are closely correlated into group ${g_{1}\dots g_{c}}$, 
$c=17$ is number of components. Features $i$ and $j$ are closely correlated if the value 
$|c_{i,j}|$ in the matrix is above a threshold, 0.98. The members of 4 out of 53 
corresponding correlation group with $|c_{i,j}|$ above 0.98 are presented in Table~\ref{tab:replicate}.

The next step is to decorrelate the features within a component, still keep one from every 
group, $g_{i}$. This process seeks to maximize the information about the class contained 
in each component and minimize the correlation within a component. There was no statistically 
significant decrease in validation accuracy when we started discarding mutually 
correlated features within a group, but the accuracy on the validation set decreases 
significantly when we discard mutually correlated features of different components. 

The third step is to select a bank of $n$ features from each component of pipeline. 
We sort ${F_{1}\dots F_{n}}$ in each component by the mutual information about the class 
\textit{benign} or \textit{suspicious}. We then use a greedy iterative selection process, 
initialized by selecting the $f_{1}$ with the highest correlation with the class. Features are 
added one per each component until the gain in accuracy is small or until a limit on the 
features $m$ is reached. This method enforces the replication of invariant features.   

\begin{table*}[]
\small
\resizebox{\textwidth}{!}{
\begin{tabular}{|c|c|c|c|}
\hline 
 group 1    & group 2   & group 3   & group 4   \\ \hline \hline
 commit.SquashedInsts   & commit.SquashedInsts  & commit.NonSpecStalls  & branchPred.condIncorrect  \\
 lsq.squashedStores  & iew.lsq.forwLoads & l2.ReadSharedReq\_misses  & commit.op\_class\_0::No\_OpClass \\
 iew.memOrderViolationEvents & dtlb.rdMisses    & rename.serializingInsts   & iew.iewExecSquashedInsts \\
 fetch.SquashCycles & iew.SquashCycles  & membus.trans\_dist::ReadSharedReq & iew.lsq.thread0.ignoredResponses \\
 iew.lsq.forwLoads  & fetch.SquashCycles    & l2.ReadSharedReq\_mshr\_miss\_latency & iq.iqSquashedInstsIssued \\
 decode.SquashCycles    & iq.SquashedNonSpecRemoved & dcache.ReadReq\_mshr\_miss\_latency   & iew.iewDispSquashedInsts \\
 iq.SquashedInstsExamined    & decode.SquashCycles  & l2.ReadSharedReq\_accesses    & branchPred.RASInCorrect \\
 lsq.squashedLoads  & iew.memOrderViolationEvents   & l2.ReadSharedReq\_miss\_latency & iq.fu\_full::FloatMemWrite \\
 iew.SquashCycles   & fetch.IcacheSquashes  & dcache.ReadReq\_mshr\_misses   & commit.op\_class\_0::FloatAdd \\
 iew.BlockCycles    & lsq.squashedLoads     & tol2bus.trans\_dist::ReadSharedReq    & fetch.PendingQuiesceStallCycles \\
 memDep.conflictingStores   & iq.SquashedInstsExamined & commit.membars & iew.lsq.thread0.rescheduledLoads \\
 dtb.rdMisses   & iew.BlockCycles   & dcache.ReadReq\_misses    & commit.branchMispredicts \\
 iq.SquashedNonSpecRemoved  & memDep.conflictingLoads   & l2.ReadSharedReq\_mshr\_misses    & branchPredindirectMispredicted \\
 rename.SquashCycles    & iq.fu\_full::MemRead  & membus.trans\_dist::ReadResp  & commit.op\_class\_0::SimdCvt \\
 memDep.conflictingLoads    & lsq.squashedStores    & rename.serializeStallCycles & iq.fu\_full::IntAlu \\
 rename.UndoneMaps  & rename.UndoneMaps    & mem\_ctrls.selfRefreshEnergy   & iew.branchMispredicts \\
 fetch.IcacheSquashes   & memDep.conflictingStores  & iq.NonSpecInstsAdded  & iew.predictedNotTakenIncorrect \\
 iq.SquashedOperandsExamined & rename.SquashCycles  & tol2bus.trans\_dist::ReadResp & tol2bus.snoop\_filter.tot\_requests \\
 \hline             
\end{tabular}%
}
\caption{Subset of highly correlated features. Features are ranked by correlations between 
the true class labels and the feature value from top to bottom and left to right. 
 Features in each group have high mutual 
correlation but belong to different pipeline components and are used to 
construct \textit{Replicated Perceptron detectors.}}
\label{tab:replicate}

\end{table*}
\begin{figure}[!htbp]
\centering
  \includegraphics[width=0.9\linewidth]{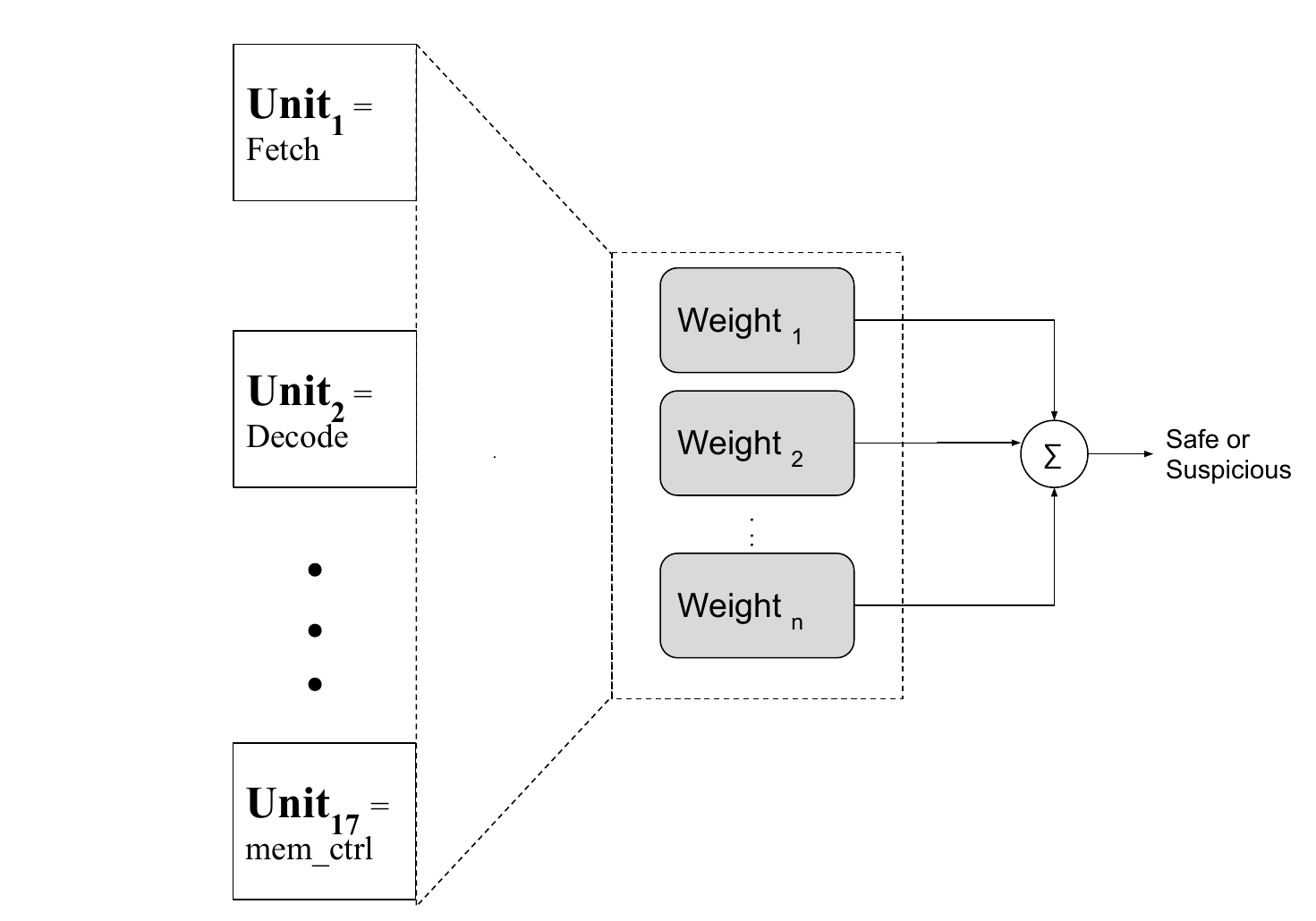}
  \caption{Architecture of PerSpectron} 
  \label{fig:predictor}
\end{figure}

\subsection{Hardware Design}
We simplify our input values into a hardware-friendly representation.  
In particular, we design a $k$-sparse binary feature to represent
values of microarchitectural statistics, where a $k$-sparse binary feature
has $k$ \texttt{1}s in the vector and \texttt{0}s elsewhere. Specifically, the $k$-sparse
binary feature vector is represented as $x \in \{0, 1\}^{u}$, where $u$ is the
total number of features and the $t^{th}$ entry $x_{t}$ is a 0/1 indicator, denoting
whether the $t^{th}$ scaled statistic is less than 0.5 or higher than 0.5. 
Scaled statistic $t$ is the value of the $t^{th}$ feature divided by the 
corresponding maximum value for a given sampling point. Maximum values are 
stored in a two-dimensional matrix $M$, with $u$ rows and $s$ columns. 
The value on the $i_{th}$ row and $j_{th}$ column is the maximum value for 
the counter $i$ at the execution point $j$. 

The example shown partially in Figure~\ref{fig:phenomm} is represented as a vector 
of inputs that when normalized to the maximum value of the counter would generate 
input vectors, which in this case are distinctive signatures. 

\textit{[f1 = ReadResp, 
f2 = commitNonSpecStalls, 
f3 = PendingQuiesceStallCycles, 
f4 = CleanEvict]}

{\small
\begin{align*}
suspicious: <0,1,0,0> \\
suspicious: <1,0,1,0> \\
suspicious: <0,0,0,1>  \\ 
safe: <1,1,0,0> 
\end{align*}
}
The above vectors show the one-hot representation and $k$-sparse
binary feature for three attacks and one safe program 
(example shown in Figure\ref{fig:phenomm}). 
We see that regardless of the order and the position of each feature, 
the $k$-sparse representation for four program are distinctive, and 
only contain 0/1 values, thereby simplifying the prediction problem. 

\subsection{Training and prediction}
We train our perceptron using the 106 features that we selected from offline training. 
The result of this training is a vector of 106 weights that will be used for building 
one final perceptron in hardware for prediction (Figure~\ref{fig:predictor}).

\subsection{Computing the perceptron output}
Since \texttt{0} and \texttt{1} are the only possible input values to the perceptron,
multiplication is not necessary to compute the dot product. Instead, we simply
add the weight when the input bit is \texttt{1} and subtract when the input bit is 
\texttt{-1}. Furthermore, only the sign bit of the result is needed to make a prediction.
This allows the other bits of the output to be computed slowly without waiting
for a final value.

\subsection{Performance Overhead}
PerSpectron has negligible performance impact since the sampling and
prediction happens in hardware and in parallel to the execution. The inference
step is not on any critical performance path for the processor. The inclusion
of any additional hardware counters not already present in the
microarchitecture is also of little concern for overhead, as
microprocessor design usually includes hundreds of counters in various parts
of the chip as a means to debug and verify processors.

Previous work adapting perceptron learning to branch prediction was done under
very tight timing constraints, requiring complex, high-speed arithmetic logic.
For example, the original perceptron branch prediction paper
suggests an expensive Wallace-tree adder~\cite{HPCA2001Jimenez} taking the
same area and energy as a large integer multiplier circuit.  However, for our
purposes the timing constraints are much more relaxed. The central computation
of perceptron-based inference, the dot-product computation, can be performed
using a very modest circuit that sequentially adds partial products. This 
generates a result in a number of cycles proportional to the number of
inputs, {\em i.e.} on the order of 100 cycles for our perceptron. Since the
inputs are binary, the circuit simply adds or subtracts weights corresponding to inputs
with the value of 1 or 0, respectively. Such a circuit would take negligible area and energy
and deliver a prediction in a very timely manner compared with the sampling
interval.

\subsection{Using PerSpectron within the System}
Once the detector identifies an attack, it is natural to ask how this information 
will be used within an end to end strategy for defending the system. Because any 
low-level machine learning detector can experience false positives, we are hesitant 
to suggest termination of a process that is suspected to be malicious. Rather, we use 
PerSpectron as a low-level detector that provides information to the system to 
trigger further mitigations. We discuss two possible classes of mitigations next.

\subsubsection{Hardware Mitigations} 
We envision our technique being deployed with the ability to update the neural 
weights using a vendor distributed patch reflecting training with the 
most recent known classes of attacks. The weights are small and few so the 
overhead of distributing and applying the patch would be modest.

The output of a perceptron before being thresholded can be used as a 
confidence measurement that can also be passed along to the operating system 
should an attack be suspected. Based on the category of the attack and 
the confidence, the technique uses a low-cost solution. For example, for 
mitigating cache attacks, the technique could apply changes to replacement policies 
to make set eviction harder or enable index randomization in the LLC like Qureshi's 
CEASER~\cite{CEASER}, or making unsafe loads invisible in the cache 
hierarchy~\cite{safespec,yan2018invisispec}
to mitigate the attack only when suspicious activity was predicted with high confidence.
For mitigating branch predictor attacks, when an attack is suspected, we can inject 
noise into the branch predictor, or invoke load fences~\cite{taram_csf19} so that it 
occasionally reverses its taken/not-taken prediction. Increasing the frequency of 
the noise increases the time for an attack to succeed and decreases performance, however.  

\subsubsection{Software Mitigations}
In order to generalize to other classes of attacks, the weights of replicated 
detectors can be adjusted to cover an arbitrary group of attacks provided 
that we are exposed to them during training. The detectors are ``software 
defined,'' meaning there is a secure update process to configure them.  
They are also replicated per pipeline component. So, we can add classifiers for 
additional attacks using the same infrastructure if new attacks emerge.  
\section{Methodology}\label{method}
We use the gem5~\cite{gem5} cycle-level simulator.
Table~\ref{table:GEM5} gives the parameters of the simulated architecture. 
We used the FANN C library~\cite{FANNnissen03} to implement PerSpectron to 
work with gem5. We used the scikit-learn python library to measure the performance 
of other machine learning models, including $K$-nearest neighbors, Decision Tree, 
Logistic Regression, and Neural Network. 

\begin{sloppypar}
\noindent\textbf{Data:}  
For speculative attacks, we simulate SpectreV1,SpectreV2, SpectreRSB, 
Meltdown, breakingKSLR (based on Meltdown) and CacheOut attack. For cache 
attacks, we run Flush+Flush, Flush+Reload, and Prime+Probe along with their 
respective profiling-calibration phases. Calibration programs threshold distinguishing 
cache hits and cache misses based on different cache attack techniques~\cite{PrimeProbe2015last,FlushReload2014Yarom, GrussFlushFlush}. 
 
For benign programs, we run individual SPEC CPU 2006 applications~\cite{spec2006}. 
The workloads include C compression programs modified to do most work in memory 
(rather than I/O), optimization scheduling, Ethernet network simulator, high-rank 
artificial intelligence programs, discrete event simulation, gene sequence protein 
analysis, the A* algorithm, and more. 
 
We have written an interface to gem5 that dumps statistics once every 10K, 50K, and 100K 
instructions, and samples all event counters for each program. Using this tool, we 
collect multi-dimensional time-series traces of applications.

\noindent \textbf{Statistics:}
We examined 1159 microarchitectural counters, including features related to 
cycle accounting and micro-operation ($\mu$op) flow, stall decomposition, 
precise memory access events, latency events, precise branch events, core memory 
access events, and other core and uncore events. 
For each event (if applicable) we measure total number, cycles, rate, average and distribution. 
For each counter, we maintain a maximum possible value for each sampling simulation point. 
Each statistic is normalized over the maximum value of that counter.

We include events related to stalls and differentiate between types of stalls and 
among various points in the pipeline. We differentiate between read and write 
latency, and what came from the master ports and was forwarded to the slave 
(requests and responses). Likewise, what came from the slave and was forwarded 
to the master. We also examined features related to energy consumption in different 
microarchitectural units and pipeline stages. Also tracked are complete power state 
machine statistics such as Active, Idle, Active Power-Down, Precharge Power-Down, 
and Self-Refresh values. 

\begin{table}[!htbp]
\small
\centering
\begin{tabular}{|c|}
\hline
\textbf{Architecture}  \\ \hline
X86 O3CPU 1 core Single Thread at 2.0GHz \\ \hline

\textbf{Core}  \\ \hline
Tournament branch predictor\\
16 RAS entries, 4096 BTB entries\\
LQEntries=32, SQEntries=32, ROBEntries=192\\
fetch/dispatch/issue/commit width=8\\
numPhysIntRegs=256,numPhysFloatRegs=256 \\ \hline

\textbf{L1 I-Cache}  \\ \hline
32KB, 64B line, 4-way \\ \hline

\textbf{L1 D-Cache}  \\ \hline
64KB, 64B line, 8-way \\ \hline

\textbf{Shared L2 cache}  \\ \hline
2MB bank, 64B line, 8-way,  \\
mshrs=20, tgtsPerMshr=12, writeBuffers=8  \\ 
tagLatency=20, dataLatency=20, responseLatency=20\\ \hline
\end{tabular}
\caption{Parameters of simulated architecture}
  \label{table:GEM5}
\end{table}

\noindent \textbf{Training and Validation:} We train the perceptrons for 1000 epochs, or 
until the training error falls below 0.4.  
We used 3-fold stratified splitting with randomization. The class split for each fold for 100K setting is as follows:  \\
Class: [benign malicious]\\
train - [4820, 1500]  | test - [2420, 740]
\end{sloppypar}
\section{Security Analysis}\label{sec:whac}\label{sec:varSpec}

We structure the security analysis of PerSpectron around three questions: 
(1) How well does PersPectron detect the different attacks? We seek to 
show that it can indeed reliably detect all the classes of attacks it 
was exposed to during training; (2) Can attackers evade detection by 
modifying the attacks? We consider different evasion strategies and discover 
that PerSpectron is robust against evasion attempts; and (3) Could 
PerSpectron generalize to detect novel cache based or speculation attacks? 
We show using examples of new attacks (that were not part of the training set) 
that PerSpectron is indeed able to detect them, likely indicating that it is 
detecting the basic signatures of these attacks in the microarchitecture. 
This is an encouraging sign that the approach can help protect against 
zero day microarchitectural attacks (novel attacks not seen before) 
even before we include them in the training set.

\subsection{Resilience to  evasion}\label{sec:robustness}
It is natural to expect that an attacker would attempt to change their attack to avoid 
detection using evasive attacks. In general, this is the problem of adversarial attacks 
against machine learning, which has been considered even against hardware 
detectors~\cite{RHMD2017}.  Although solutions such as RHMD can also be applied to our 
detector, we believe that the nature of microarchitectural attacks makes it difficult 
for attackers to evade detection without interfering with the attack, meaning that the 
attack features are strong discriminators that cannot be hidden without disabling the attack.

To support this hypothesis, we investigate the robustness of PerSpectron to two evasion 
techniques in this section: (1) Resilience to polymporphic evasion (where the attacker 
attempts to produce different binaries implementing the attack); and (2) Resilience to 
bandwidth reduction mimicry: a common evasion strategy is to slow down the rate of the 
attack to weaken the microarchitectural signal. 

\subsubsection{Polymorphic evasion}
We generated variants of the attack PoC code-snippets used in this study to investigate 
whether our predictor is resilient to such variations. This is a typical strategy 
used by malware to evade signature based 
detectors~\cite{PaulKocher,paulKocherSpectreAttacks}. PerSpectron was not exposed 
to any of the modified attacks: they were not used in any of the feature selection 
phases or final model training. We considered the following transformations:
\begin{compactitem}
\item Moving the leak to a function that cannot be inlined. 
\item Add a left shift by one on the index.
\item Use \texttt{x} as the initial value in a \texttt{for()} loop.	
\item Check the bounds with an \texttt{AND} mask, rather than \texttt{<}.
\item Compare against the last-known good value.
\item Use a separate value to communicate the safety check.
\item Leak a comparison result in which the attacker is assumed to provide both \texttt{x} and \texttt{k}. 
\item Make the index the sum of two input parameters.
\item Do the safety check into an inline function.
\item Invert the low bits of \texttt{x}.
\item Use \texttt{memcmp()} to read the memory for the leak.
\item Pass a pointer to the length.
\end{compactitem}

\begin{figure}[!htbp]
\centering
 \includegraphics[width=\linewidth]{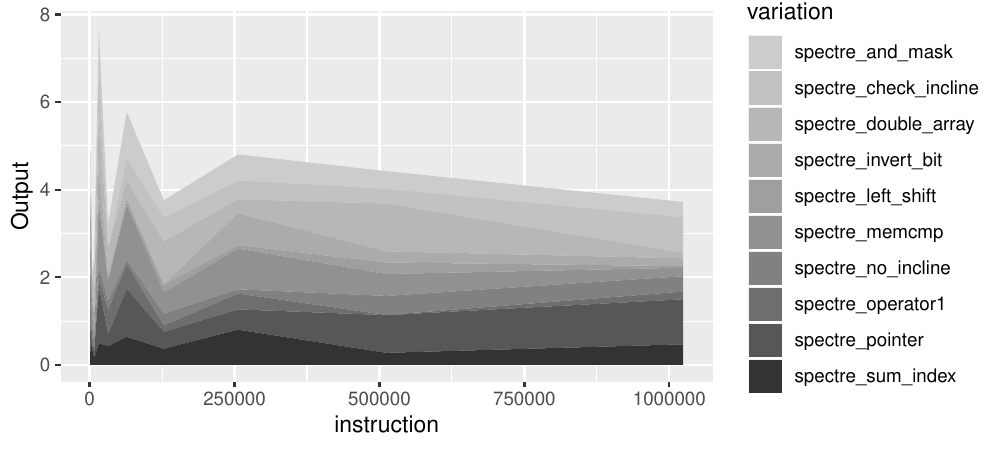}
 \caption{Perceptron output vs. number of instructions for 
 Spectre variations with same leakage frequency. All variants
 were flagged as suspicious at the same sampling interval. }
 \label{fig:vars}
\end{figure}

Figure~\ref{fig:vars} shows a stacked line chart of perceptron output
pre-threshold versus the number of instructions simulated for different
versions of a Spectre attack with the same leakage frequency. All variations
were detected and triggered the perceptron output as suspicious at the same
sampling interval, showing that the variations do not cause failure in the
detection.

\subsubsection{Bandwidth reduction evasion}
A typical microarchitectural attack encompasses a priming phase used
to place the system into an initial state, {\em e.g.} flushing the cache, a
speculation phase that exposes a victim's data, a disclosure gadget
phase that readies the data for transmission, and a final disclosure
primitive phase that leaks the data to the adversary. We seek to
detect suspicious activities before the disclosure phase
begins so that appropriate countermeasures are launched on time to
prevent leaking the first byte of data.

\begin{figure}[!htbp]
\centering
  \includegraphics[width=0.9\linewidth]{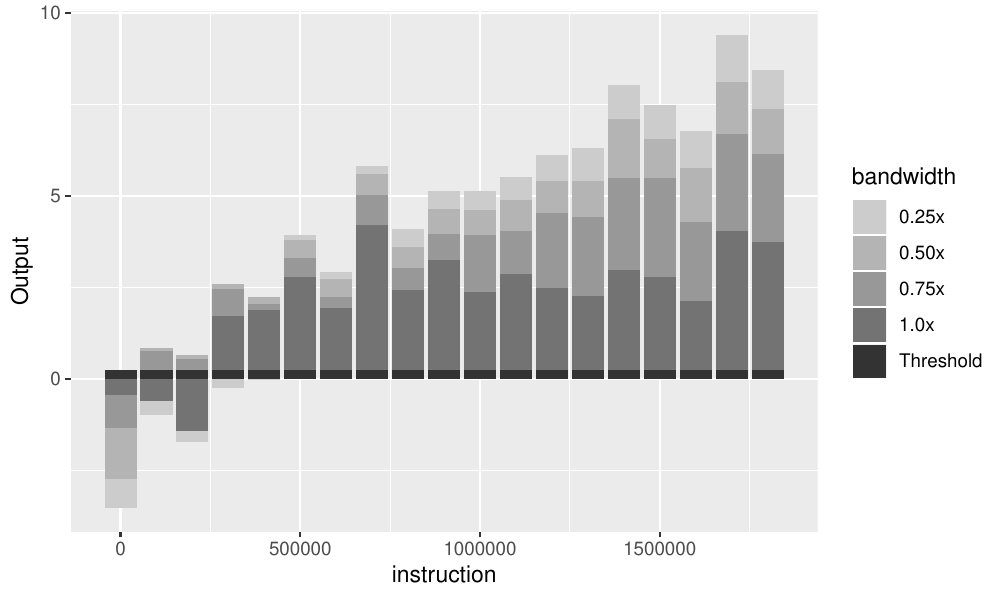}
 \caption{Perceptron output pre-threshold vs. number of instructions for various SpectreV1 bandwidths.}
 \label{fig:band}
\end{figure}

Figure~\ref{fig:band} shows perceptron output pre-threshold versus the
number of instructions simulated for various SpectreV1 bandwidths.
We reduced the bandwidth of Spectre by 75\%, 50\%, and 25\% over the
original attack code by injecting safe source code into the program.
The safe code was injected after the disclosure primitive phase as
well as before the priming phase, in proportion to the ratio of
bandwidth reduction. Our safe code does not tamper with branch history 
or the secret cache line.

Figure~\ref{fig:band} shows that even though the unmodified Spectre
(1.0x) saturates the output faster, all lower bandwidth versions stay
above the cutoff threshold after the first complete phase of the
attack, showing that PerSpectron detects the suspicious
activity. Further reducing the bandwidth to less than 25\% leads to no leakage
of data. While PerSpectron continues to detect an attack at lower bandwidth, 
data is not leaked even after many tries within 2,000,000,000 executed instructions.

Bandwidth evasion is based on timing the completion of all attack atomic 
tasks to fit within the sampling interval. It is a problem with the 
low sampling frequency of the software detector. Li and 
Gaudiot~\cite{Gaudiot2020} demonstrated that performance counter-based 
detectors can be evaded by changing the bandwidth of the attack so that 
it runs inside the 100 ms sampling interval of the detector. They identified 
three "Atomic Tasks" that, if interrupted, will disable the attack: 
(1) Flushing cache lines - 10 $\mu$s
(2) Mistraining branch predictor - 13 $\mu$s
(3) Attempting to infer the secret byte that is loaded into cache - 38 $\mu$s. 
The authors concluded that putting the attack to sleep after all three tasks were 
completed is the optimum evasion strategy for an attack.  This means that the 
attack runs for 61 $\mu$s before being put to sleep, which allows Spectre to run 
inside the 100 ms sampling interval of the detector. Our sampling interval is 
3 $\mu$s, which gives 20 sampling intervals within the 61 $\mu$s run time it takes 
to complete all three tasks, making PerSpectron resistant to this evasion strategy.  
The authors acknowledged that future work should be done on a dedicated hardware 
detector to reduce performance overhead.
Therefore evasion is made more difficult by decreasing the sampling 
interval to below the run time of essential tasks of the attack.

\subsection{Generalization to Other Attacks}
When evaluating trained models at development time, our goal was to estimate 
the generalization accuracy, {\em i.e.} the predictive performance of a model on future 
(unseen) attacks. Thus, we use $K$-fold cross-validation (CV) to characterize 
the distribution of possible model behaviors on workloads not present in the 
training set. 

We start by splitting our dataset into three parts: a training set 
for model fitting, a validation set for model selection, and a test set for 
the final evaluation of the selected model. We merge the training and validation sets 
(excluding the test set) after model selection. At every fold, we remove all the 
samples belonging to attacks in the test set and hence they were not used for 
model selection. Therefore, our cross validation confidence interval of 
(0.9979$\pm$0.0065) simulates new data that the model has not seen before.   
Thus, we are naturally evaluating the generalization of the detection to attacks 
not part of the training or validation set.

We also explored whether PerSpectron can generalize to newly discovered attacks 
within the category of microarchitectural and speculative execution attacks. 
Specifically, we consider the CacheOut attack, which was disclosed after 
PerSpectron was developed~\cite{cacheOut}. 
CacheOut attacks Read from Line Fill Buffer, a feature that PerSpectron 
identified as invariant prior to disclosure of CacheOut. CacheOut 
(94\% true-positive detection) demonstrates PerSpectron’s robustness on 
detecting advanced MDS attacks.
We also held out SpectreV2 from any of the independent training set of 
3-cross-validation and were able to detect it (91\% true-positive) 
providing evidence of robustness in detecting new variations. 

Both CacheOut and SpectreV2 transmit values via a Flush+Reload channel. 
To assure that PerSpectron is detecting both recovery and transmission 
portion of attacks, we used different cache channels for the attacks in 
the test set and the training set of the each CV fold (see Table~\ref{table:CV}). 

Note that PerSpectron does not rely on a cache channel for detection; it relies on replicated detectors and invariant features that monitor contention in all CPU components i.e., buffers and ports. Our CV results confirm that there are distinct signatures for these attacks that should hold across their different variations, and by extension to newly discovered ones, which shows our detector to be generalized to more variations.  

\begin{table*}[!h]

\centering
\small
\resizebox{\textwidth}{!}{%
\begin{tabular}{|c|c|c|}
\hline
k'th test fold & $D_{k}$ & $D_{-k}$ \\ \hline
1 & spectreRSB, spectreV2, cacheOut, breakingKSLR, prime+probe & meltdown, spectreV1, flush+flush, flush+reload \\ \hline
2 & spectreV1, spectreV2, cacheOut, flush+reload & meltdown, spectreRSB, breakingKSLR,  flush+flush, prime+probe \\ \hline
3 & spectreV2, cacheOut, meltdown, breakingKSLR,  flush+flush & spectreV1, spectreRSB, flush+reload, prime+probe \\ \hline
\end{tabular}%
}
\vspace*{-1mm}
\caption{Estimating the risk using cross validation. At each fold, we excluded one version of each categories of attacks, trying to predict the attack using a model trained on data that does not contain that attack.}
\label{table:CV}
\end{table*}

\subsection{Attack Coverage}
Table~\ref{tab:compare} shows that DT-CART~\cite{Quinlan86inductionof} with 
MAP~\cite{Malware2015} features was not able to detect Prime+Probe and SpectreV1 
with a bandwidth smaller than 0.75x post leakage. DT-CART using PerSpectron features 
was able to detect Prime+Probe but still was not able to detect attacks with 
bandwidth lower than ($<0.75x)$. KNN was unable to detect calibration-ff, calibration-pp, 
and Prime+Probe even after leakage. Interestingly KNN could detect polymorphic 
attacks using PerSpectron features. Logistic Regression with MAP features suffered 
from high false positives, and could not detect Prime+Probe and polymorphic 
attacks until post leakage. Perceptron, with the large feature set of 1159, had 
false negatives on Prime+Probe pre-leakage. A neural network with MAP features 
suffered from false positives on memory and interrupt intensive workloads when 
they were dropped out of the training set. Neural network with PerSpectron features 
had fewer false positives and could detect 11 out of 12 polymorphic attacks. 
{\em PerSpectron} could detect all of the attacks before they leaked with only 
false negatives on initial and pre-leakage samples. This includes the 
Prime+Probe as well as Flush+Flush 
attacks\footnote{Note that Perspectron does not detect occurrence of \textit{leakage}. 
PerSpectron detects suspicious activity related to microarchitectural attacks 
that leads to leakage. Some variations don't leak in the simulation 
environment.}.
\section{Experimental Evaluation}\label{results}
This section details our experimental results exploring different aspects 
of the behavior of PerSpectron.  Specifically, we first study the 
impact of the sampling granularity, discovering that prediction accuracy 
improves up to a sampling rate of 10K instructions.  We also compare 
PerSpectron to other models and feature sets and see that it achieves 
substantially higher accuracy than these models for the types of attacks 
we consider. Finally, we explore the individual feature behavior across 
the pipeline to demonstrate both the interpretability of the behavior of 
the detector and how the signatures manifest in different pipeline stages, 
supporting a hypothesis that PerSpectron is both generalizable and 
difficult to evade.

\subsection{Sampling granularity and threshold}
\begin{figure}[!htbp]
\centering
 \includegraphics[width=0.75\linewidth]{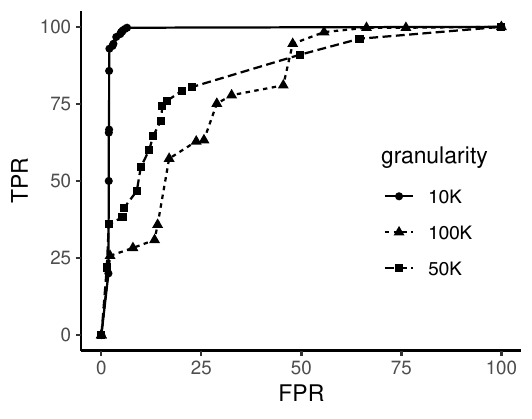}
 \caption{ROC curve for different sampling granularities. 
 A 10K interval is better than 50K and 100K intervals. }
 \label{fig:rocG}
\end{figure}

We tested different sampling granularities and threshold values.  We
present the receiver operating characteristic (ROC) curves in
Figure~\ref{fig:rocG} to show the relationship between true
positive and false positive rate. The figure shows ROC curves under
three different sampling intervals: 10K, 50K and 100K instructions.
We can see the 10K interval is better than the 50K and 100K.

We found that a threshold of 0.25 is the best with an area under the curve.
PerSpectron has high sensitivity and specificity (AUC of 0.9949) to atomic 
tasks and benign code. The prediction statistics are as follows: mean 
accuracy is 0.9979, standard deviation is 0.0015 and accuracy range 
of (0.9914, 1.0000).

\subsection{Comparison with other models and feature sets}\label{compareMLs}
To compare different models' performance, we specifically used standard K-fold 
cross-validation (CV). Standard CV is the unbiased estimate of prediction error, 
which is the expected loss on future instances. More precisely, our CV is defined as 
follows. Let there be $N=|D|$ data cases in the training set. Denote the data in 
the $k$'th test fold by $D_{k}$ and all the other data by $D_{-k}$ (Table~\ref{table:CV}). 
At every fold, we remove all samples belonging to attacks and benign programs in the 
test set so they are not used for training the model, analytically removing the effect 
of the $i$'th training case.
Benign programs on these fold are chosen so the class proportions are roughly 
equal in each fold. 

Our experiments with multi-way classification achieves a near-perfect F1-score on 
the training set. However, due to the limited number of attacks per category, excluding 
test and training set for K-fold-CV was impractical for multi-way classification. 
Thus, in this section we report the accuracy of binary classification comparing 
with previous works. Table~\ref{table:CV} shows which attacks were chosen for 
training and which attacks are excluded.\footnote{We excluded CacheOut from all 
the folds to further stress the CV results---especially to compare the 
robustness of models on detecting advanced MDS attacks.}

We investigated the features used in previous works that had success using 
Logistic Regression for detecting malware. Table~\ref{tab:compare} shows 
Logistic Regression trained with features similar to MAP~\cite{Malware2015}, achieving 
75.5\% accuracy in detecting microarchitectural attacks while majority labeling 
yielded 74.4\% accuracy. These results show that the signatures of microarchitectural 
attacks are different than signatures of malware attacks. In addition, memory, 
branch, and interrupt intensive benign workloads such as h264ref, povray, gcc, 
sjeng, and gobmk were shown as false positive using the features from previous work.  
We also tried Logistic Regression using the 106 highly correlated features we 
extracted via the PerSpectron algorithm. The accuracy increased to 89.7\%, 
showing that using a highly correlated feature set improves detection accuracy. 
The K-nearest neighbor algorithm showed a high mean accuracy of 94.8\% with the 
best $k=3$ and PerSpectron features. However, KNN is not suitable for implementation 
in hardware due to its high overhead and classification latency.

Table~\ref{tab:compare} shows that the decision tree with CART~\cite{Quinlan86inductionof}
algorithm has 87.1\% accuracy with MAP~\cite{Malware2015} features and 
90.5\% accuracy with the 106 highly correlated PerSpectron features. Decision tree, 
on the other hand, leads to hard decisions in different regions of input space, 
causing poor performance on classifying attacks that were unseen during the 
training. Lastly, CART was unable to correctly classify low bandwidth attacks. 

We also investigated neural networks using the features from previous works~\cite{Malware2015}, 
yielding a low accuracy of 80.2\%. We noticed that the values of features kept constant 
during the different phases of attacks show that the previous work did not use 
features suitable for microarchitectural attack detection.
The neural network, when trained with 106 features selected with the PerSpectron 
algorithm, yields an accuracy of 98.2\%. However, a deep neural network has 
high hardware overhead and classification latency, making it unsuitable for 
implementation in hardware. 
    
Finally, PerSpectron gave the highest accuracy with 99.7\% and a tight confidence 
interval. PerSpectron does not suffer from a greediness problem similar to decision 
trees. Unlike the single perceptron, miss-classification by one component can be 
recovered by the replicated feature detector from the other component of the pipeline. 

\begin{table*}[!h]
\centering
\begin{adjustbox}{width=\textwidth}
\small
  \begin{tabular}{|c|c|c|c|c|c|c|c||c|}
  \hline
  \textbf{Model} & \textbf{DT-CART*} & \textbf{DT-CART} & \textbf{Logistic Regression*} &
  \textbf{Perceptron} & \textbf{KNN} & \textbf{NN*} & \textbf{NN} & \textbf{PerSpectron} \\ 
  \hline \hline 
  \textbf{Feature} & \textbf{MAP} & \textbf{PerSpectron} & \textbf{MAP} & \textbf{1159 features} &
  \textbf{PerSpectron} & \textbf{MAP} & \textbf{PerSpectron} & \textbf{PerSpectron} \\ 
  \hline \hline
  Mean Accuracy & 0.8718 & 0.9058 &  0.7594 & 0.8974 & 0.9487 & 0.8026 & 0.9822 & 0.9979 \\ \hline
  95.00\% Confidence  & 0.8718 $\pm$ 0.1005 & 0.9058 $\pm$ 0.0120 & 0.7594 $\pm$ 0.0018 & 
  0.8974 $\pm$ 0.2030 & 0.9487 $\pm$ 0.1435 &  0.8026 $\pm$ 0.0026 &
  0.9822 $\pm$ 0.0089  &  0.9979 $\pm$ 0.0065\\ \hline

  \makecell{False Positives\\ $>20 samples$ }   &  \makecell{ dealII,gcc,\\gobmk,bzip2}  & 
  dealII, gcc & \makecell{h264ref,povray,gcc\\sjeng,gobmk\\}  & \makecell{gobmk,dealII\\}  &
  \makecell{sjeng,gobmk \\}  & \makecell{ dealII,povray\\bzip2,gobmk} &  \makecell{gobmk,dealII\\}  &
  \makecell{gobmk\\ }  \\ \hline
  \makecell{False Negatives\\$post/preleakage$} & 
  \makecell{ prime+probe,\\spectre$<0.75x$\\post leakage} & 
  \makecell{spectre$<0.75x$,\\polymorphic\\post leakage} &  
  \makecell{prime+probe,\\spectre$<0.75x$,\\polymorphic\\post leakage} & \makecell{prime+probe,\\spectre$<0.75x$\\pre leakage} &  
  \makecell{calibration-ff/pp*\\ post leakage}  &  
  \makecell{prime+probe\\spectre$<0.75x$\\pre leakage}  &  
  not post leakage &  
  not post leakage \\ \hline
  Hardware Complexity & low & low & low  & low &  high & high &  high & low \\ \hline
	\end{tabular}
\end{adjustbox}
\raggedright{\begin{small}
\textit{*pp: prime+probe}\end{small}}
\vspace*{-3mm}
\caption{ML Model and Feature Set comparison}
\label{tab:compare}
\end{table*}

\subsection{Interpretation through feature analysis}\label{analyse}
An important aspect of our model is its interpretability. In perceptron based detectors, 
when features in an active program carry high positive weights, suspicious activity 
is indicated, whereas a negative weight indicates a benign program. Thus, the weights 
attached to each feature potentially convey insightful information related to how 
each feature correlates to the output giving us a detailed understanding of the 
footprint of these microarchitectural attacks in our feature space. This 
interpretability contrasts with deep neural network models whose features are 
automatically generated from the data, and therefore resist interpretability. 
In the following, we take advantage of this visibility to explore the footprint 
of the attacks in groups of our features related to different components of the 
architecture.

\noindent \textbf{Rename unit features:}
We found that \textit{CommittedMaps} is an invariant feature that contributes highly 
to different classes of the suspicious activity including speculative and cache 
attacks. The Committed Map table holds the speculative mappings from ISA registers 
to physical registers. If enabled, this allows a single-cycle reset of the pipeline 
during flushes and exceptions of any kind. On every branch, the Rename Map tables are 
used to allow single-cycle recovery on a branch misprediction. Otherwise, pipeline 
flushes require multiple cycles to ``unwind'' the ROB to write back in the rename 
state at the commit point, one ROB row per cycle. This functionality explains why 
\textit{CommittedMaps} is invariant to different versions of transient execution 
attacks. Polymorphic evasions that uses contention on the Register File were 
detected preleakage after including \textit{CommittedMaps} features even by the 
simplest models {\em i.e.} (DT-CART) which shows this feature is highly informative.  

The other invariant features from the rename stage are  \textit{serializeStallCycles}, 
\textit{LQFullEvents} and \textit{tempSerializingInsts}. We also observe that 
Flush+Flush-type attacks have an unusually high number of temporary serializing
instructions in the rename unit which also manifest in the commit unit as a 
high number of stalls.

\noindent \textbf{Memory controller features}: 
One feature that we found to be both strongly correlated to attacks and 
invariant against a range of them is the \textit{bytesReadWrQ} selected from 
Memory Controller. The \textit{bytesReadWrQ} tracks the number of DRAM read 
requests that are serviced by the write queue. We observe that the access rate 
to the write queue is abnormally high in all attacks early on before the 
completion of the first iteration of attacks. Including this feature reduces 
the false positive for memory intensive workloads. This feature illustrates 
the fact that most attacks attempt to read data freshly evicted from the 
caches while constructing a covert channel inside the tight speculative 
window at a rate much higher than typical even in memory intensive workloads. 

Another two invariant features from \textit{mem\_ctrl} were \textit{bytesPerActivate}
with large positive weights, and \textit{wrPerTurnAround} with lowest negative weight.
Feature \textit{bytesPerActivate} is the number of accessed bytes per row activation in 
DRAM and \textit{wrPerTurnAround} is the number of writes before turning the bus 
around for reads. We noticed that including these two features reduces the number 
of false positives and remain discriminatory for reduced bandwidth attacks in all 
the models. 

Another invariant feature is the \textit{selfRefreshEnergy} from \textit{mem\_ctrl}. 
The system's DRAM will enter a low power self-refresh mode when it is not being 
actively utilized. Many variations cause high volumes of memory accesses to 
non-cache resident arrays (for example, due to Flush operations common in Flush+X attack 
variants). Including \textit{selfRefreshEnergy} decreases the false positive rate 
by 2\% and makes the predictor more resilient to polymorphic evasions\footnote{This 
is an interesting observation because flush operations are also essential for Rowhammer 
attacks~\cite{rowhammer}. We have not simulated the Rowhammer attack due to the 
limitations of simulator, but this result is promising with respect to the generalization 
of PerSpectron to other micro-architectural attacks that share concepts with 
the attacks we considered in our feature set.}.

\noindent \textbf{L2 bus features}: 
We found that  features \textit{ReadCleanReq}, \textit{WritebackClean} and 
\textit{readsharedreq} remain discriminative across a range of attacks and 
evasive variations.  Essentially, these events capture cache evictions, but 
do not focus on the mechanism (whether using Flush or cache priming to cause 
the eviction of a cache line). Our results show previous detections were not able 
to detect prime+probe correctly without causing an increase in false positives. The 
reason is prime+probe behaves similar to the large memory footprint applications when 
monitored by high-level features. We were able to detect prime+probe by including 
\textit{ReadCleanReq}, \textit{readsharedreq} and  \textit{WritebackClean} without 
causing an increase in false positives. 

\noindent \textbf{Fetch unit features}:
The fetch unit contains highly correlated features that are invariant to variations 
of stalls, flush, miss and traps used in the polymorphic attacks. The reason is 
because stalling a pipeline involves stopping not only the pipe stage experiencing 
{\em i.e.} the data dependency, but previous stages as well. And therefore such signals 
propagated backward to the fetch stage. Features \textit{PendingQuiesceStallCycles}, 
\textit{IcacheSquashes}, \textit{MiscStallCycles} and \textit{PendingTrapStallCycles} 
are mutually decorrelated in the fetch unit but they have high correlation with stalls 
and traps in other components, {\em i.e.} \textit{commit.NonSpecStalls}, 
\textit{lsq.thread0.rescheduledLoads} and \textit{dcache.blocked:no\_mshrs}, 
constructing a perfect set for replicated detectors as explained in Section~\ref{alg}.
This combination particularly enabled us to classify orthogonal speculative attack 
techniques~\cite{paulKocherSpectreAttacks} ({\em e.g.}, speculative read in spectre vs. 
page fault trap in Meltdown) with perceptron in a linearly separable space. 

We believe that the interpretability of our features provides confidence that 
they are strong discriminators that are difficult to evade. It also provides 
confidence that the large number of features across the whole pipeline capture 
different aspects of attack signatures, supporting generalization across 
different attacks, provided they manifest in one or more of the components of 
the feature space.
\section{Related Work}\label{related}
We discuss the related work in two categories: (1) Detection-based solutions which 
attempt to monitor behavior and identify attacks, and (2) other defenses against 
microarchitectural attacks. We focus primarily on solutions that are rooted in 
hardware and do not discuss software-only solutions.
\subsection{Detection based solutions}
Prior work~\cite{cacheBasedDetection2016Chiappetta,
CloudRadar2016,
BlackHatFogh,
payer2016hexpads,
ICCAD2015Wang,
Duppel2013Zhang, 
mushtaq:cel-01824512} has used hardware performance counters to detect cache-based 
side-channel attacks. These approaches are based on the assumption that all cache 
attacks cause unusual cache access patterns. To detect the attacks, they use cache 
related counters such as L1miss, L3hit, L1D and L1I accesses and others. These 
counters are highly biased toward the cache access patterns and are unable to 
detect attacks like Flush+Flush, which do not access the cache. It would be easy for 
an attacker to evade such approaches since they use only a few simple counters. 

FortuneTeller~\cite{gulmezoglu2019fortuneteller} uses Recurrent Neural
Networks and hardware performance counters at the OS-level to predict
Meltdown, Spectre, RowHammer and ZombieLoad.  They tested 36 counters with
their selection method. They used 3 counters to profile for anomaly detection,
as the desktop processors Intel Core i5 and i7 are limited to 3 programmable
counters. The three selected counters are \textit{L1InstMiss},
\textit{L1InstHit}, \textit{LLCMiss}.  The overall performance degradation of
FortuneTeller is 24.88\% for 10$\mu$s sampling
rate~\cite{gulmezoglu2019fortuneteller}.

Similar to our approach, prior work has explore the idea of using a hardware detector 
to detect malicious software~\cite{Malware2015,kazdagli-16}. Unlike PerSpectron, 
these classifiers target general malware, which are software attacks with a footprint 
that should manifest in the committed instructions of the program. In contrast, 
our goal is to detect microarchitectural attacks that are inherently exploiting 
hardware. As a result, the signature of these attacks is reflected in the behavior 
of the processor pipeline overall, and not only in the features present in the committed 
stage of the pipeline. Thus, we focus on features related directly to these microarchitectural 
attacks and view our solution to be solving a different problem and looking at different 
feature spaces. CC-Hunter~\cite{chen2014cc} detects contention-based CPU side channels by 
detecting unsual contention patterns. In a similar vein, ReplayConfusion~\cite{yan2016replayconfusion} 
records a running application, then replays it on a different cache configuration 
to detect whether contention patterns are sensitive to the configuration of the 
cache (indicating an attack). GPUGuard applies a similar strategy of detecting 
microarchitectural attacks using a decision tree classifier and then invoking defenses.
Sadly, it is limited to the context of GPUs, where these attacks are still limited~\cite{xu-19}. 

While some studies have explored some high level features such as cache hits/misses and 
branch missprediction, they do not consider individualized features in the pipeline stage. 
PerSpectron's feature space is much more detailed because signature of microarchitectural 
attacks is more complex and different than the signature of malware.

\subsection{Other Defenses}
A number of solutions have been proposed to protect against micro-architectural attacks 
including side- and covert-channels on caches, as well as transient execution attacks 
such as Spectre and Meltdown.

\subsubsection{Defenses against Cache-based Side channels}
In this section, we review proposed work on defending against side channel attacks on
shared cache hierarchies. NoMo~\cite{domnitser2012non} proposes dynamic cache partitioning; 
it controls the number of ways for each application to interfere with the attacker's 
ability to prime the cache. CATalyst~\cite{liu2016catalyst} is another partitioning technique 
that partitions the cache into security sensitive and a non-secure regions. 
RPCache~\cite{wang2007new} and CEASER~\cite{qureshi2018ceaser} try to prevent the attacker 
from evicting target lines from caches by randomizing the mapping of addresses to cache sets.
TimeWarp\cite{martin2012timewarp} and FuzzyTime\cite{hu1992reducing} add noise to the system 
clock to prevent attacks that measure cache access latency and execution time.
RIC\cite{kayaalp2017ric} exploits the inclusion property of the LLC to prevent LLC side-channel 
attacks. This approach prevents the attacker from replacing the victim's data from the local 
core caches, thus preventing the leakage of secret data.

\subsubsection{Defenses against transient execution attacks}
To defend against SpectreV1, Intel and AMD proposed to use serializing 
instructions~\cite{amd,arm_css} near the branch, eliminating the performance 
gain of speculation execution. A low overhead hardware solution to mitigate 
SpectreV1 attacks is Context Sensitive Fencing~\cite{taram_csf19}. In this approach, 
serializing instructions are injected dynamically to the target of the branches 
based on the run time conditions or taint analysis engine.

To mitigate Spectre-BTB and SpectreRSB attacks, Intel and AMD~\cite{amd,intelanalysis} 
include new instructions (i.e, IBRS) in that control speculation around indirect branches. 
Experiments show that these instructions impose high overhead. Retpoline~\cite{retpoline} 
is a software mitigation technique against a Spectre-BTB attack that converts indirect 
branches to a sequence of direct call and return and uses an infinite loop or a 
serializing instruction to prevent the CPU from speculating on the target of an indirect branches.

Several hardware-based techniques are proposed to mitigate different types of speculative 
execution attacks. DAWG\cite{kiriansky2018dawg} propose partitioning the cache at the cache 
way granularity to provide isolation between protection domains, which can prevent leakage 
through cache-based side channel attacks. SafeSpec~\cite{safespec}, 
InvisiSpec~\cite{yan2018invisispec} and CleanupSpec~\cite{saileshwar2019cleanupspec} 
hide transient microarchitectural side effects from the committed state of the processor,
unless they are committed. NDA\cite{weisse2019nda} and SpecShield\cite{barber2019specshield} 
prevent the propagation of potential secrets to the speculative instructions that may leak 
the secrets by forming a side channel. STT~\cite{yu2019speculative} propose using taint 
tracking to prevent the forwarding of the sensitive data to speculative instructions. 

SpecCFI~\cite{koruyeh2019speccfi} proposed to use hardware-level control-flow integrity to 
reduce the range of possible indirect call targets to a small set, and then protecting this 
small set of call sites using the known lfence-based technique to mitigate all 8 variants of 
Spectre. In contrast to these efforts, PersPectron uses a different approach of detecting 
the footprint of these microarchitectural attacks, enabling rapid reaction to them, rather 
than on re-architecting the processor to interfere with these attacks.
 
We should note that several of the detection and defense mechanisms discussed either impose 
high performance overhead to the system and/or only mitigate a specific variant of the attacks. 
We believe that PerSpectron offers an inexpensive, always-on solution that can provide wide 
coverage against microarchitectural attacks. We also believe that there are interesting 
possibilities when combining PerSpectron with other defenses, for example to raise the 
alarm to activate other defenses to avoid their overhead unless an attack is suspected.

\section{Conclusion and Future Work}\label{future}
This paper proposes a detection mechanism for a broad range of micro-architecture 
side-channel attacks in hardware. Previous approaches achieve low accuracy in detecting 
unseen variations. Software-only methods are also prone to bandwidth evasion.
PerSpectron design uses two techniques from Vision (in ML); \textit{replicated detectors} 
and \textit{invariant feature selection}, leveraging a large set of features available 
in hardware from all components of the processor including ports, buffers, buses, and uncore.

PerSpectron mapped a non-linearly separable problem to linear, which was then 
separable by perceptron and is implementable in hardware. The result is competitive 
with a more complex deepNN that is not easily implementable in hardware. The use of machine 
learning gives an opportunity to cover other attacks in the future by expanding the 
feature set and updating the perceptron weights.

Many features used in PerSpectron are invariant to the nature of the attack and are 
difficult to evade: for example, speculation attacks exploit mispredictions or exceptions. 
The large number of features potentially make prior evasion attacks prohibitively 
expensive~\cite{RHMD2017}. In the future, we hope to evaluate and, if necessary improve the 
security of PerSpectron against adversarial ML attacks, for example by exploring approaches 
that randomize the feature subset being used over time similar to the RHMD defense~\cite{RHMD2017}.

\section*{Acknowledgment}
We thank Jeffrey N. Collins for his help and comments during the drafting of
this paper.  This research was supported by National Science Foundation grants
CNS-1938064, CCF-1912617, and generous gifts from Intel Corporation.

%
%


%
\end{document}